\documentclass[review]{elsarticle}

\usepackage{amsfonts}
\usepackage{amsmath}
\usepackage{lineno,hyperref}
\modulolinenumbers[5]
\usepackage{graphicx}
\usepackage{epstopdf}
\usepackage{algorithmic}
\usepackage{subcaption}
\usepackage{nicefrac}
\usepackage{units}
\usepackage[acronyms]{glossaries}

\journal{Elsevier}

\makeatletter
\providecommand{\doi}[1]{%
  \begingroup
  \let\bibinfo\@secondoftwo
  \urlstyle{rm}%
  \href{http://dx.doi.org/#1}{%
    doi:\discretionary{}{}{}%
    \nolinkurl{#1}%
  }%
  \endgroup
}
\makeatother

\newacronym{dns}{DNS}{direct numerical simulations}
\newacronym{cfd}{CFD}{computational fluid dynamics}
\newacronym{pod}{POD}{proper orthogonal decomposition}
\newacronym{gpr}{GPR}{Gaussian process regression}
\newacronym{gp}{GP}{Gaussian processes}
\newacronym{gpu}{GPU}{graphics processing unit}
\makeglossaries
\glsdisablehyper

\begin{document}
\begin{frontmatter}

\title{Data recovery in computational fluid dynamics through deep image priors}

\author[main]{M.~T.~Henry de Frahan\corref{cor1}}
\ead{marc.henrydefrahan@nrel.gov}

\author[main]{R.~W.~Grout}
\ead{ray.grout@nrel.gov}

\cortext[cor1]{Corresponding author}

\address[main]{High Performance Algorithms and Complex Fluids, Computational Science Center, National Renewable Energy Laboratory, 15013 Denver W Pkwy, ESIF301, Golden, CO 80401, USA}

\begin{abstract}

  One of the challenges encountered by computational simulations at
  exascale is the reliability of simulations in the face of hardware
  and software faults. These faults, expected to increase with the
  complexity of the computational systems, will lead to the loss of
  simulation data and simulation failure and are currently addressed
  through a checkpoint-restart paradigm. Focusing specifically on
  \gls{cfd} simulations, this work proposes a method that uses a deep
  convolutional neural network to recover simulation data. This data
  recovery method (i) is agnostic to the flow configuration and
  geometry, (ii) does not require extensive training data, and (iii) is
  accurate for very different physical flows. Results indicate that
  the use of deep image priors for data recovery is more accurate than
  standard recovery techniques, such as the \gls{gpr}, also known as
  Kriging. Data recovery is performed for two canonical fluid flows:
  laminar flow around a cylinder and homogeneous isotropic
  turbulence. For data recovery of the laminar flow around a cylinder,
  results indicate similar performance between the proposed method and
  \gls{gpr} across a wide range of mask sizes. For homogeneous
  isotropic turbulence, data recovery through the deep convolutional
  neural network exhibits an error in relevant turbulent quantities
  approximately three times smaller than that for the
  \gls{gpr}. Forward simulations using recovered data illustrate that
  the enstrophy decay is captured within $10\%$ using the deep
  convolutional neural network approach. Although demonstrated
  specifically for data recovery of fluid flows, this technique can be
  used in a wide range of applications, including particle image
  velocimetry, visualization, and computational simulations of
  physical processes beyond the Navier-Stokes equations.

\end{abstract}


\end{frontmatter}

\glsresetall

\section{Introduction}
As modern computational efforts reach exascale, hardware and software
faults will increasingly cause difficulty in completing
simulations~\cite{Brown2010}. Current research in hardware systems and
software frameworks~\cite{Gropp2004, Hoemmen2011, Teranishi2014,
  Cappello2014, Gamell2015, Grout2017} continues to develop techniques
to detect and anticipate system failures; however, assuming that a
fault has been detected and signaled, the data loss from these
failures will require data recovery processes~\cite{Lee2015}. This
work addresses this challenge. Current computational codes rely
on a checkpoint and restart paradigm to recover from faults, requiring
either significant memory consumption for frequent checkpoints or
large resimulation efforts. The ability to recover the missing data
without resorting to data checkpoints has the potential to increase
simulation resilience.

In the context of \gls{cfd}, data recovery has been explored using a
variety of machine learning approaches. Gappy \gls{pod} has shown
particular success in reconstructing missing
data~\cite{Everson1995,Tan2003,Venturi2004}. This approach relies on
combining \gls{pod} with least-squares estimates~\cite{Yates1933,
  Little2002} and data from \acrlong{dns} snapshots. Venturi and
Karniadakis~\cite{Venturi2004} expanded on the methods proposed by
Everson and Sirovich~\cite{Everson1995} and used this technique to
reconstruct missing data of unsteady flow past a cylinder. Other
approaches rely on \gls{gpr}, often referred to as Kriging in
geophysics, which is a a reconstruction technique that uses the mean and
covariance of the \acrlong{gp} prior. The prior's covariance is determined
by a kernel whose hyperparameters are optimized using training
data. Gunes et al.~\cite{Gunes2006} compare \gls{pod}-based and
\gls{gpr}-based solution reconstruction and show that \gls{gpr}
interpolations are particularly effective for unsteady flows
(including instability regions), whereas \gls{pod}-based methods are
advantageous when the temporal resolution is high. In addition to
these methods, Lee et al.~\cite{Lee2015} proposed a ``resimulation''
method in which the missing data region is resimulated using
appropriate initial and boundary conditions. This new method is
evaluated for lid-driven cavity flows and flows past a cylinder at low
Reynolds numbers. Lee et al.~\cite{Lee2017} combined the gap-tooth
algorithm, previously developed for dynamic systems~\cite{Gear2003},
multiresolution information fusion, and auxiliary data to construct a
general framework for fault-tolerant \gls{cfd}. The method is
demonstrated to work well for simulations of the heat equation and
lid-driven cavity flow.

Recently, the deep learning community has been proposing methods for
data recovery in the field of inverse image reconstruction problems,
which include denoising, inpainting, and
super-resolution,~\cite{Goodfellow2014,Burger2012,Dosovitskiy2015,Lefkimmiatis2016,Ledig2017,Tai2017,Lai2017}. Deep
convolutional neural networks, particularly generative adversarial
networks, have been very successful at solving this class of
problems~\cite{Goodfellow2014}. Inpainting is of particular relevance
to our objective of reconstructing flow solutions. The objective of
inpainting is to fill in missing portions of a damaged image such that
the result is indistinguishable from the original image. Various
generative adversarial neural networks have been proposed for image
inpainting with notable
success~\cite{Yeh2016,Denton2016,Pathak2016,Li2017a,Sasaki2017}. Generally
speaking, these approaches have relied on training deep neural
networks with an extensive and large data set of images such that the
network learns image priors that it can use in other
configurations. The deep learning methodology used in this work was
first developed by Ulyanov et al.~\cite{Ulyanov2017} for various
inverse image reconstruction problems, including inpainting. In
contrast with previous image reconstruction solution with deep neural
networks, Ulyanov et al.~\cite{Ulyanov2017} showed that ``contrary to
the belief that learning is necessary for building good image priors,
a great deal of image statistics are captured by the structure of a
convolutional image generator independent of learning.''  Instead of
training the neural networks with a large database of images, the
authors use untrained neural networks to fit single degraded images,
using the network weights as parameters for solving the image
reconstruction problem.

In this work, we use deep convolutional neural networks, such as those
proposed by Ulyanov et al.~\cite{Ulyanov2017}, for spatial reconstruction of the flow
solution for simulations wherein some type of fault led to loss of
data, e.g.,\,processor failure. In contrast to gappy
\gls{pod}~\cite{Venturi2004}, we assume that the current gappy data are
the only available data for the reconstruction procedure. This
assumption is relevant to large simulations where it is
computationally expensive to reload data residing on the file system
and the reconstruction process is restricted to data already in
memory. One advantage of using deep convolutional neural networks is
that this approach avoids eigenmode decompositions for solution
reconstruction, which could restrict the applicability or translation of
the method to new configurations. As illustrated in this work, the
method proposed here is not specific to the flow configuration and
does not require multiple training data samples.

This paper is organized as follows. In Section\,\ref{sec:formulation}, we
present the problem formulation and define the objective function for
the data recovery problem. In Section\,\ref{sec:nn}, we detail the
architecture of the deep convolutional neural network used to perform
the data recovery process for fluid flows. In Section\,\ref{sec:results}, we
present our results by evaluating the neural network's ability to
perform data recovery for two canonical flows: laminar flow past a
cylinder, Section\,\ref{sec:cylinder}, and homogeneous isotropic turbulence,
Section\,\ref{sec:hit}. These results are compared with data recovery
performed through \gls{gpr}. Finally, conclusions and future work are
presented in Section\,\ref{sec:ccl}.

\section{Problem formulation}\label{sec:formulation}

In this work, we evaluate the performance of deep convolutional neural
networks for data recovery in \gls{cfd}. Deep convolutional neural
networks have shown particular success for solving the image
reconstruction problem~\cite{Goodfellow2014}. Image reconstruction is
analogous to data recovery because they share a similar objective to
provide synthetic data that closely match the missing data. The
image reconstruction problem can be cast as an optimization problem:
\begin{align}
  \min_x{E(x;x_0) + R(x)}
\end{align}
where $x$ is the original image that needs to be recovered, $x_0$ is
the corrupted image, $E(x;x_0)$ is the task-dependent data term, and
$R(x)$ is the image prior. In the case of inpainting, the
task-dependent data term is:
\begin{align}
  E(x;x_0) = || (x-x_0) \circ m||^2
\end{align}
where $\circ$ is the Hadamart product, $m \in \{0,1\}^{h\times w}$
represents the binary mask, and $h$ and $w$ are the image height and
width. The image prior is usually captured through the training of
convolutional neural networks using a large image database. In the
approach proposed by Ulyanov et al.~\cite{Ulyanov2017}, $R(x)$ is replaced by a
parameterization such that the optimization problem becomes:
\begin{align}
  \min_\theta{E(f_\theta(z); x_0)} \label{equ:loss}
\end{align}
where $f$ represents the convolutional neural network with parameters
$\theta$ that is initialized randomly, and $z$ is a fixed input. The
fixed input for the neural network can take many forms but is usually
chosen to be random uniform noise or smoothly varying data.
Note that the neural network input is \textit{fixed}. Given a
deteriorated image, the neural network effectively learns, by
backpropagation and network parameter tuning, the encoding necessary
to map the fixed input to an output, i.e.,\,the \textit{recovered}
image, which minimizes the loss function,~Equation\,\ref{equ:loss}.

We emphasize that physical constraints are not explicitly included in
the data recovery process. This has the advantage of enabling a
reconstruction technique that does not depend on the physical nature
of the problem. Higher fidelity can be achieved, however, by
incorporating physical constraints, as suggested in~\cite{Raissi2017,
  Sirignano2018}. The work presented here focuses on two-dimensional
reconstruction, though there is no inherent methodological limitation
to reconstructing three-dimensional data directly.

\section{Neural network architecture}\label{sec:nn}

\begin{figure}[!tbp]%
  \centering%
  \begin{subfigure}[t]{1\textwidth}%
    \includegraphics[width=\textwidth]{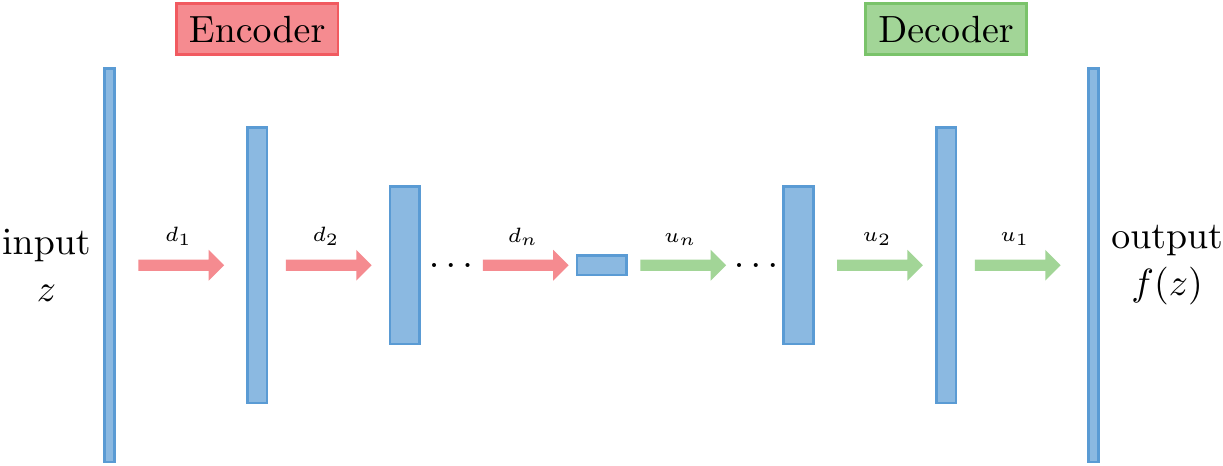}%
    \caption{Encoder-decoder architecture.}\label{fig:encoder_decoder}%
  \end{subfigure}\\[0.3cm]%
  \begin{subfigure}[b]{0.33\textwidth}%
    \includegraphics[width=0.92\textwidth]{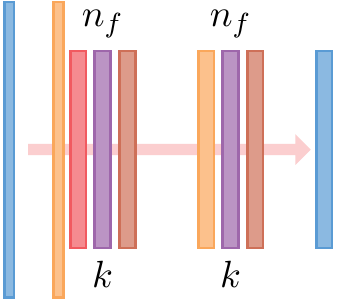}%
    \caption{Downsampling unit.}\label{fig:downsample}%
  \end{subfigure}%
  \hfill%
  \begin{subfigure}[b]{0.33\textwidth}%
    \includegraphics[width=\textwidth]{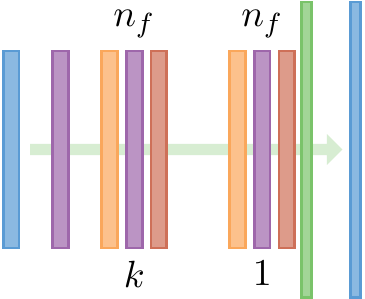}%
    \caption{Upsampling unit.}\label{fig:upsample}%
  \end{subfigure}%
  \hfill%
  \begin{subfigure}[b]{0.15\textwidth}%
    \includegraphics[width=\textwidth]{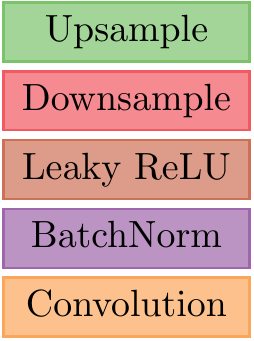}%
    \vspace*{0.4cm}%
    \caption{Legend.}\label{fig:legend}%
  \end{subfigure}%
  \caption{Deep convolutional neural network used for data recovery.}\label{fig:arch}%
\end{figure}

The network chosen for this work is a convolutional neural network
that exhibits an encoder-decoder architecture with approximately 2
million tunable parameters and no skip connections,
Figure\,\ref{fig:encoder_decoder}. This architecture enables the
network to encode the input in the latent space and then decode the
latent space representation into the reconstructed image. The
nonlinear activation function used in the network is LeakyReLU
\cite{He2015}. Downsampling was performed through simple striding in
the convolution procedure, Figure\,\ref{fig:downsample}, and
upsampling was done through nearest-neighbor upsampling,
Figure\,\ref{fig:upsample}. The number of filters in the downsampling
and upsampling layers, $n_f$, was kept fixed at 128, and the kernel
size, $k$, was fixed at 3. Experiments showed that using a fixed
smoothly varying input $z$ for the neural network imposes a smoothness
prior, which is beneficial for data recovery for fluid flows. The
optimization process was performed using Adam~\cite{Kingma2014}. The
implementation was done in PyTorch~\cite{Paszke2017}, and the learning
process was computed on a Tesla V100 \acrlong{gpu}. The number of
iterations for all the experiments was 2000, thereby reducing
the loss function by three orders of magnitude.

\section{Results}\label{sec:results}

To demonstrate the efficacy of using deep neural networks for data
recovery, we investigate two types of flows: laminar flow over a
cylinder for data recovery of large flow scales and homogeneous
isotropic turbulence for data recovery of flows spanning a wide range
of scales. We compare the deep convolutional neural network results
with \gls{gpr}. The sample points used to train the \gls{gpr} are
located in the region surrounding the mask with a depth of 10 cells,
similar to~\cite{Lee2015}. Beyond a depth of 10 cells surrounding
the masked regions, training the regressor becomes computationally
intractable because the \gls{gpr} complexity is $\mathcal{O}(n^3)$, where
$n$ is the number of training points. A radial basis function is used
as the \gls{gpr} kernel.

\subsection{Laminar flow around a cylinder}\label{sec:cylinder}
The first numerical tests of the data recovery process are performed
for the laminar flow around a cylinder ($Re = 200$). The simulation is
performed using Nalu-Wind, a low Mach Navier-Stokes solver leveraging
the Trilinos
libraries\footnote{\url{https://github.com/Exawind/nalu-wind}}; and the
$t=234\,\unit{s}$ snapshot is used for the numerical tests, at which
time the vortices behind the cylinder were fully developed. Masks
simulating data loss because of processor failure are generated in the
cylinder wake. To capture typical domain decomposition methods for
structured grids, the masks are square boxes and vary in size
depending on the number of processors used for the simulation. The
mask box length, $L_m$, ranged from $0.5D$ to $5D$, where $D$ is the
cylinder diameter, and the masks are located at 40 random locations in
the cylinder wake, leading to 240 unique masks to be applied to the
simulation data. For the reconstruction process, reflection padding is
used for the boundary conditions.

\begin{figure}[!tbp]%
  \centering%
  \begin{subfigure}[t]{0.45\textwidth}%
    \includegraphics[width=\textwidth]{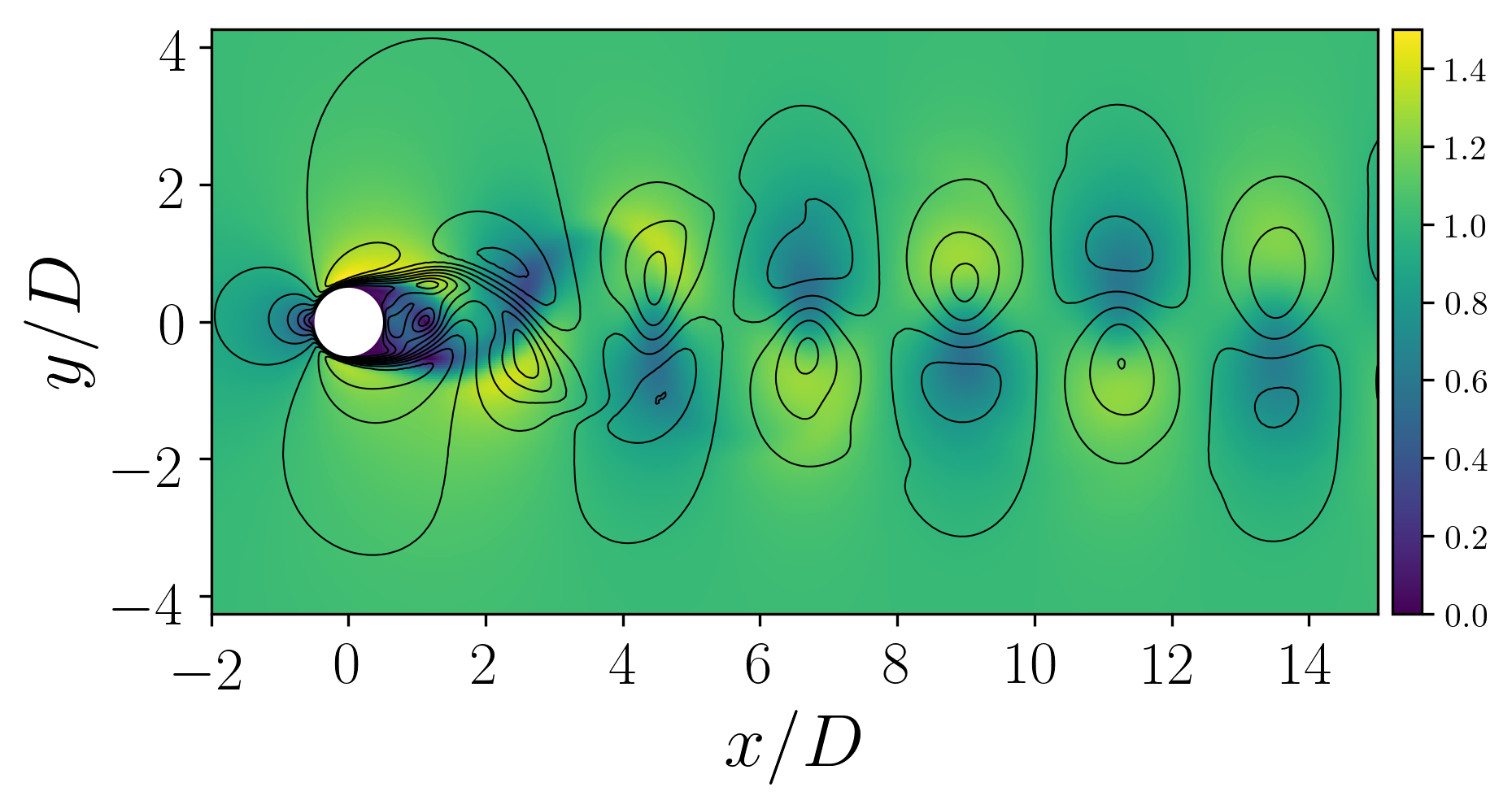}%
    \caption{Original data.}\label{fig:cyl_umag0}%
  \end{subfigure}%
  \hfill%
  \begin{subfigure}[t]{0.45\textwidth}%
    \includegraphics[width=\textwidth]{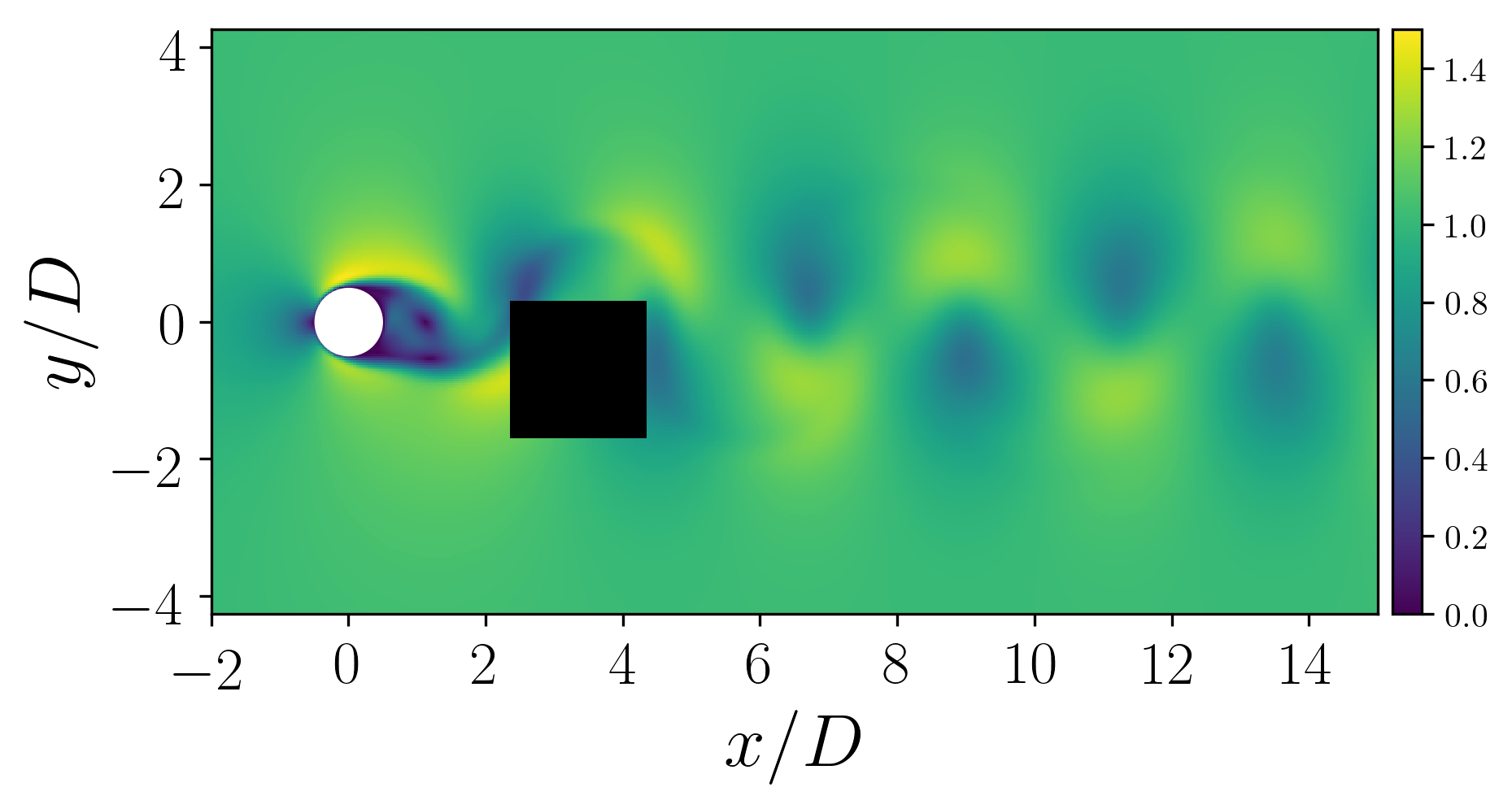}%
    \caption{Deteriorated data.}\label{fig:cyl_umag0_masked}%
  \end{subfigure}\\%
  \begin{subfigure}[t]{0.45\textwidth}%
    \includegraphics[width=\textwidth]{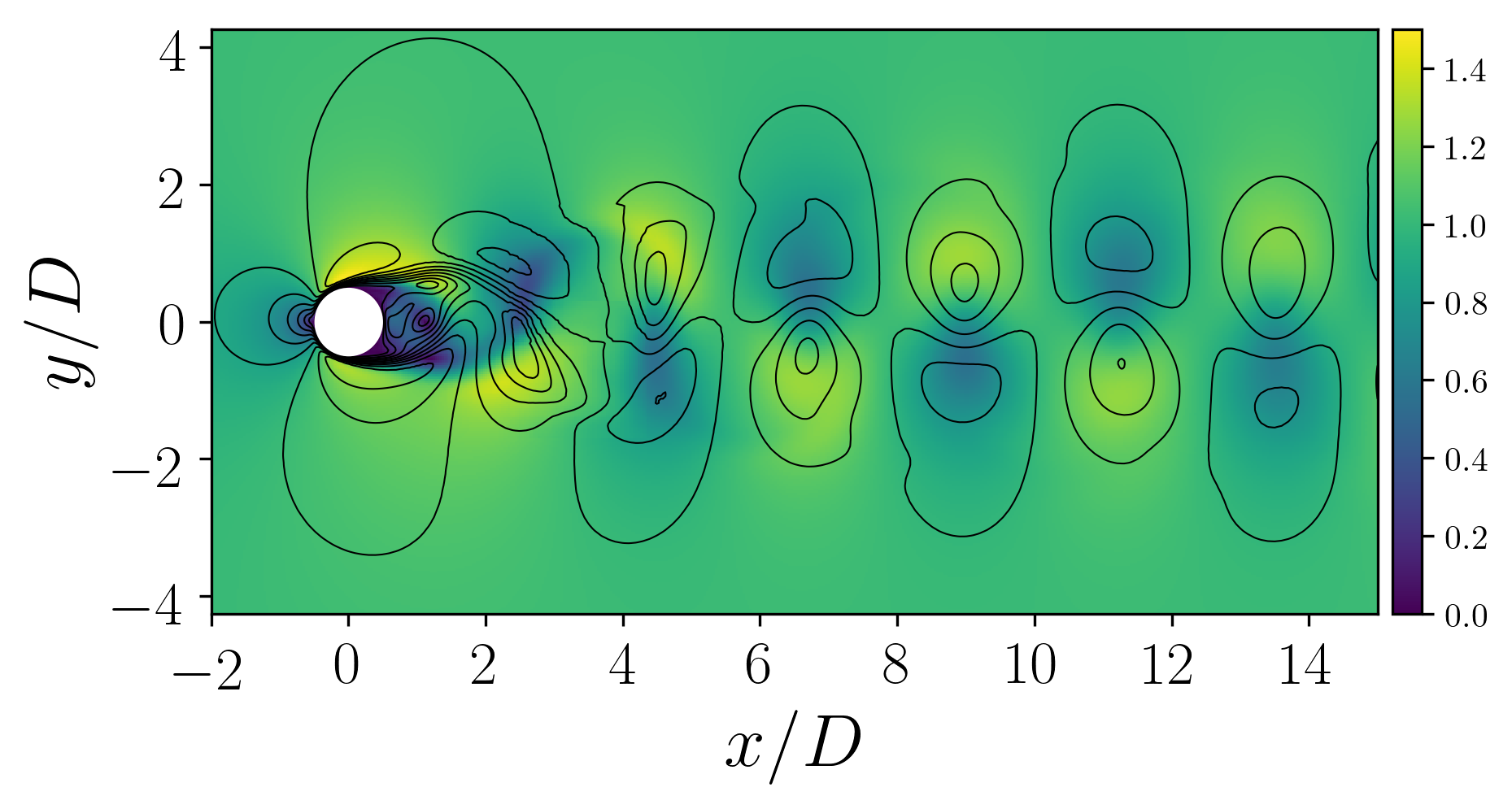}%
    \caption{Deep convolutional neural network.}\label{fig:cyl_umagr}%
  \end{subfigure}%
  \hfill%
  \begin{subfigure}[t]{0.45\textwidth}%
    \includegraphics[width=\textwidth]{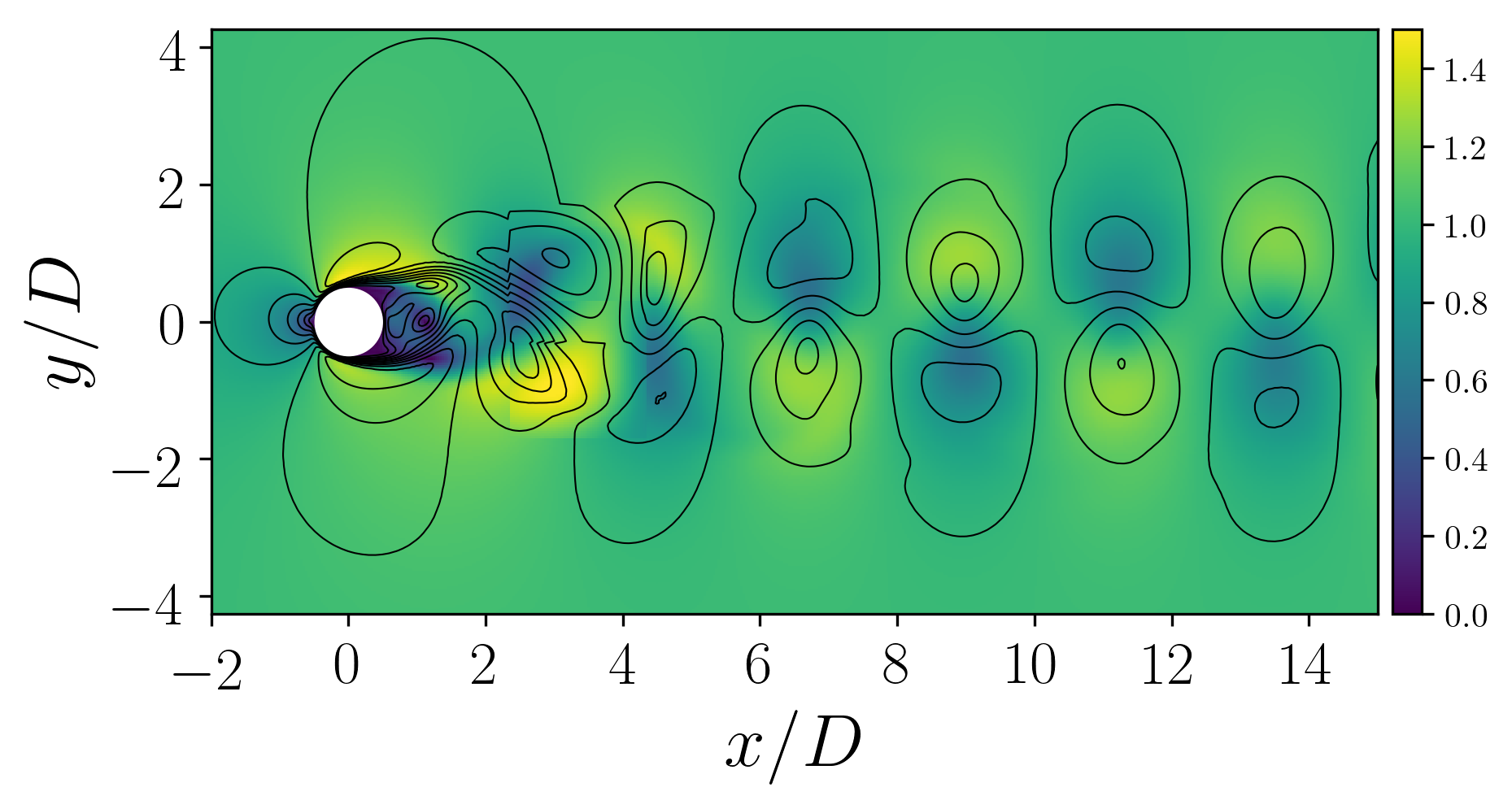}%
    \caption{\gls{gpr}.}\label{fig:cyl_umagi}%
  \end{subfigure}%
  \caption{Velocity magnitude for laminar flow around a cylinder where the mask box length is twice the cylinder diameter, $L_m=2D$.}\label{fig:illustrated_cyl}%
\end{figure}

\begin{figure}[!tbp]%
  \centering%
  \begin{subfigure}[t]{0.48\textwidth}%
    \includegraphics[width=\textwidth]{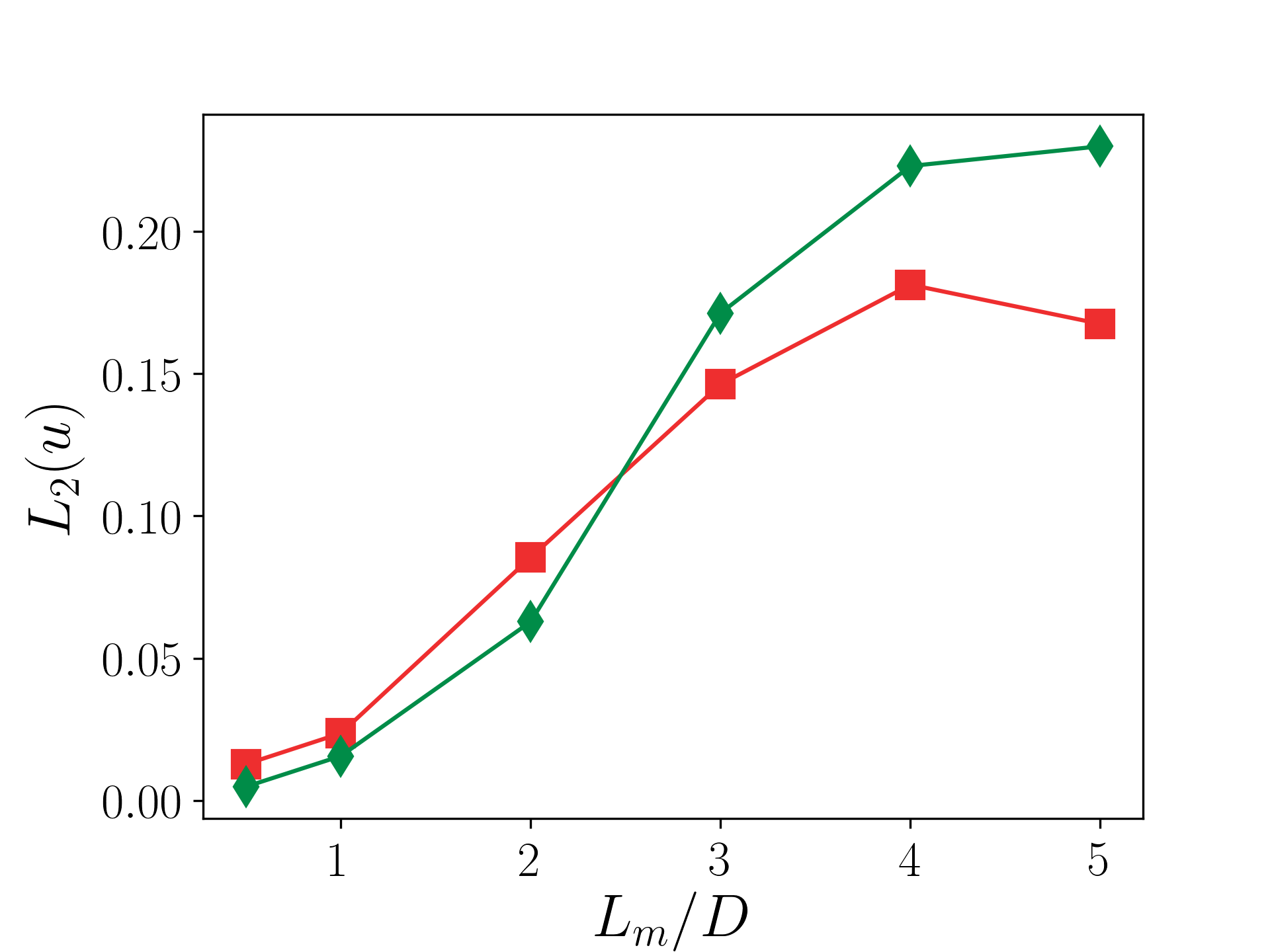}%
    \caption{$x$-direction velocity.}\label{fig:cyl_error_u}%
  \end{subfigure}%
  \hfill%
  \begin{subfigure}[t]{0.48\textwidth}%
    \includegraphics[width=\textwidth]{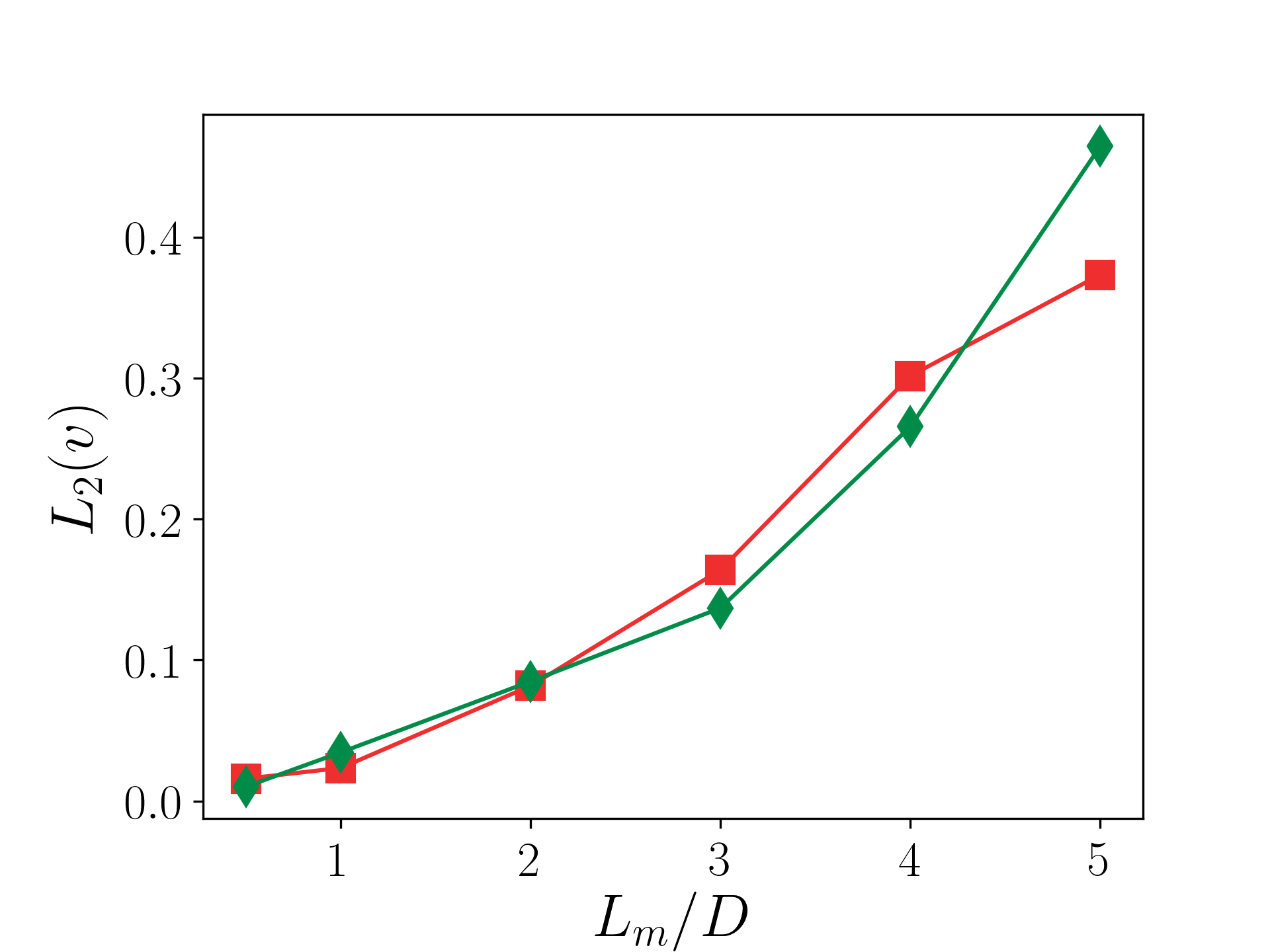}%
    \caption{$y$-direction velocity.}\label{fig:cyl_error_v}%
  \end{subfigure}%
  \caption{Average $L_2$ error norm as a function of mask box length,
    $L_m$, for laminar flow past a cylinder ($Re=200$). Red squares:
    deep convolutional neural network; green diamonds:
    \gls{gpr}.}\label{fig:cyl_error}%
\end{figure}

An example reconstruction is presented in
Figure\,\ref{fig:illustrated_cyl} for $L_m=2D$, where the deep
convolutional neural network presents a slightly better reconstructed
velocity field than \gls{gpr}. Specifically, the partially masked
vortex is more accurately reconstructed using the deep convolutional
neural network. The average $L_2$ error norm for the velocity fields
as a function of $L_m$ is presented in
Figure\,\ref{fig:cyl_error}. Both reconstruction techniques, \gls{gpr}
and the deep convolutional neural network, present similar error
profiles. At higher $L_m$ the neural network performs slightly better
than \gls{gpr} for the $x$-direction velocity, whereas it performs
similarly for all other lengths. Given the structured nature of the
flow field, it is unsurprising that \gls{gpr} performs well at
moderate mask sizes, given previously published
results~\cite{Gunes2006}.

\subsection{Homogeneous isotropic turbulence}\label{sec:hit}

For these numerical tests, we use two-dimensional slices of
homogeneous isotropic turbulence with a Taylor microscale Reynolds
number $Re_\lambda= \nicefrac{\rho_0 u' \lambda}{\mu}=133$, where
$\rho_0$ is the reference density,
$u' = \sqrt{\nicefrac{\overline{u_i u_i}}{3}}$ is the initial mean
fluctuating velocity,
$\lambda = \nicefrac{\overline{u_1^2}}{\overline{\left(\frac{\partial
        u_1}{\partial x_1}\right)^2}}$ is the Taylor microscale, and
$\mu$ is the dynamic viscosity; a turbulent Mach number
$M_t = \nicefrac{u_0}{c_s} = 0.1$, where $c_s$ is the speed of sound;
and a Prandtl number $Pr = \nicefrac{\mu c_p}{k} = 0.71$, where $c_p$
is the heat capacity at constant pressure, and $k$ is the thermal
conductivity. The reference temperature and pressure are 300~K and 1
atmosphere and the ideal gas equation of state is used to relate the
thermodynamic quantities. The domain ranges from $[0,2\pi]$ with
periodic boundary conditions. In this work, we use
PeleC\footnote{\url{https://github.com/AMReX-Combustion/PeleC}}, an
explicit compressible Navier-Stokes flow solver based on the AMReX
library\footnote{\url{https://amrex-codes.github.io}}, to demonstrate
the data recovery process. For the reconstruction process, periodic,
i.e.,\,wrapped, padding was used for the boundary conditions.

\begin{figure}[!tbp]%
  \centering%
  \begin{subfigure}[t]{0.32\textwidth}%
    \includegraphics[width=\textwidth]{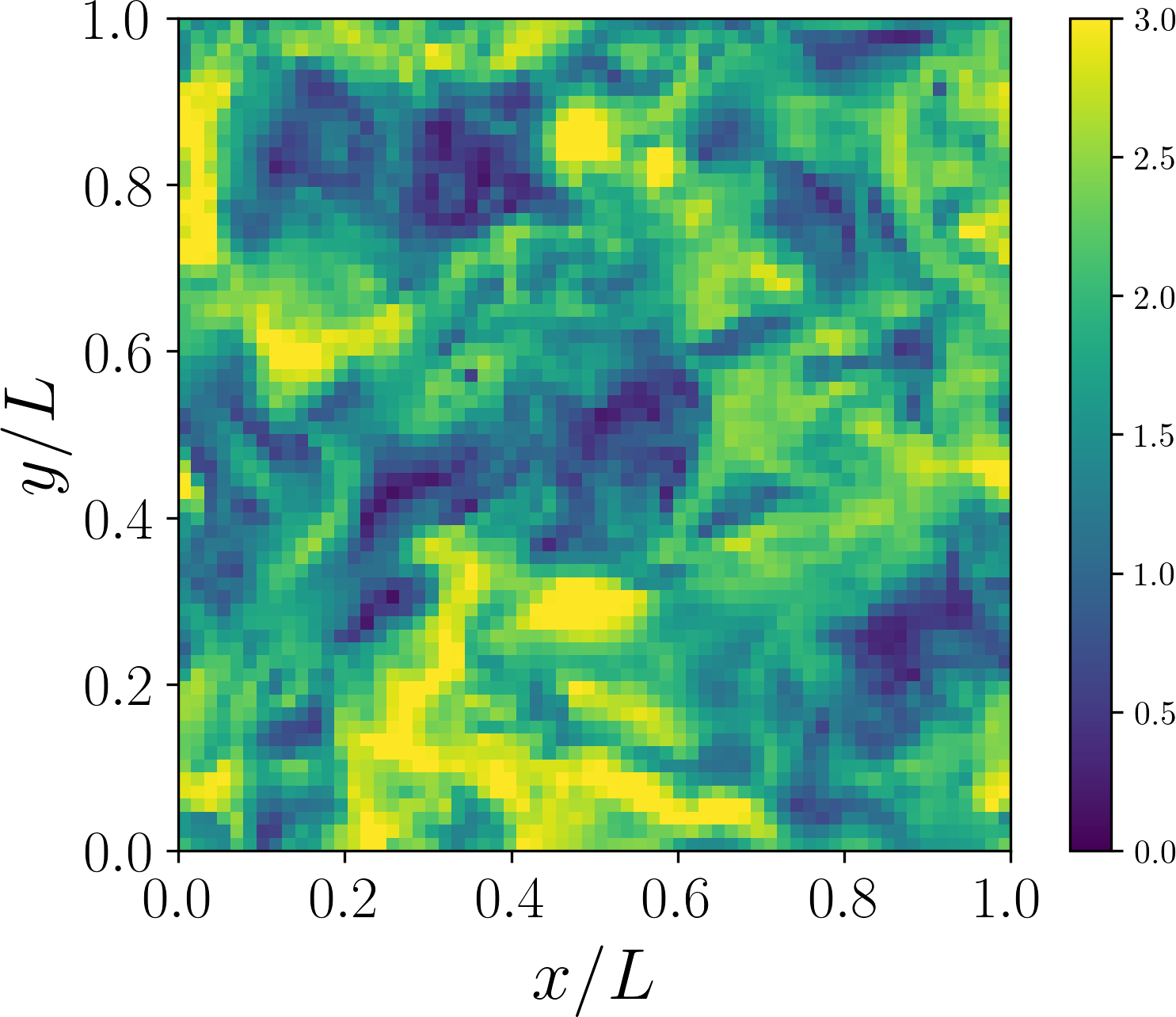}%
    \caption{Original.}\label{fig:original}%
  \end{subfigure}%
  \hfill%
  \begin{subfigure}[t]{0.32\textwidth}%
    \includegraphics[width=\textwidth]{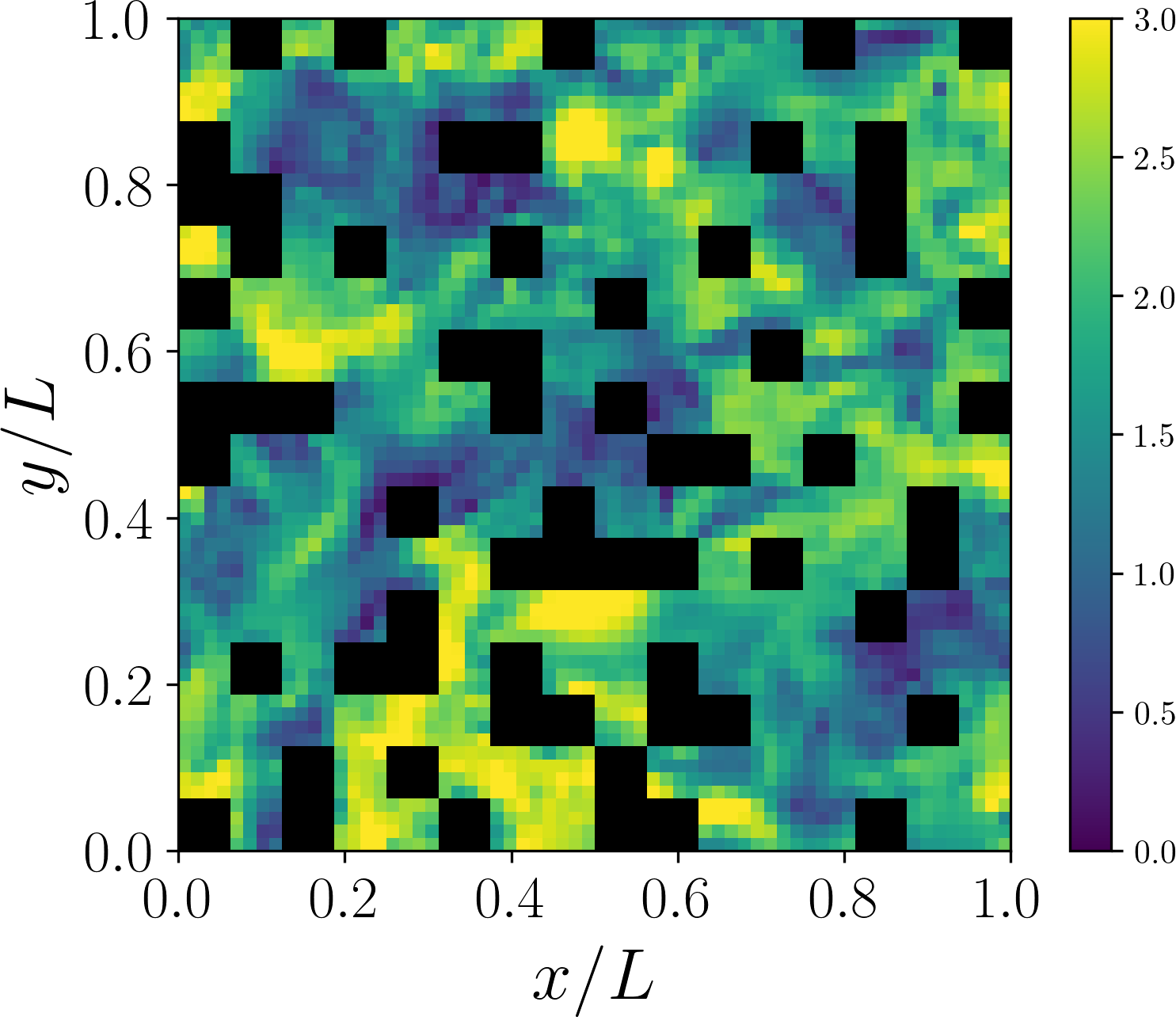}%
    \caption{Deteriorated.}\label{fig:masked}%
  \end{subfigure}%
  \hfill%
  \begin{subfigure}[t]{0.32\textwidth}%
    \includegraphics[width=\textwidth]{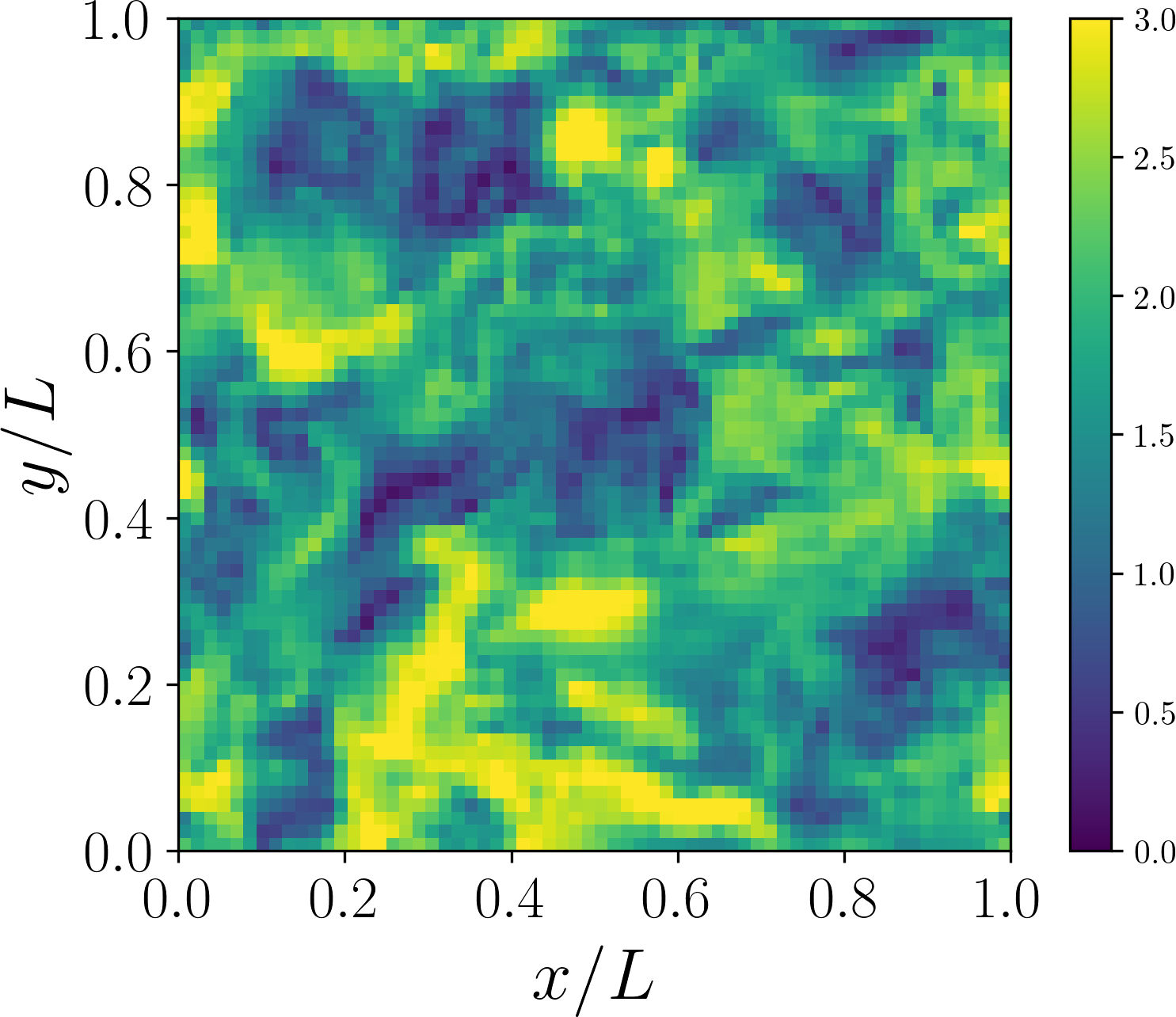}%
    \caption{Recovered.}\label{fig:result}%
  \end{subfigure}%
  \caption{Velocity magnitude in homogeneous isotropic turbulence
    illustrating the data recovery process, where 25\% of the original
    data is missing and the length scale associated to each missing
    block is 6.25\% of the domain
    ($1.5\lambda$).}\label{fig:illustrated_hit}%
\end{figure}

Initial two-dimensional data slices are generated by slicing in each
direction a numerical simulation of homogeneous isotropic turbulence
at a resolution of 64 cells in each direction, leading to 192 unique
slices. The velocities in each direction are assigned an input channel
for the neural network. Masks simulating data loss because of processor
failure are generated independently, following typical domain
decomposition. We explore two different parameters associated with the
mask generation process. The first is the total percentage of missing
data, $f$, ranging from $6.25\%$ to $25\%$. The second is the length
scale associated with each block of missing data, $L_m$, ranging from
$3.125\%$ to $50\%$ of the domain length, or $0.74\lambda$ to
$11.87\lambda$. For each pair of parameters $f$ and $L_m$, we randomly
generate ten different masks, resulting in 130 unique masks. These are
randomly applied to 100 initial slices, resulting in 1300 slices
requiring reconstruction. For each of these deteriorated slices, the
neural network parameters are tuned to optimize the reconstruction
loss function, Equation\,(\ref{equ:loss}). The data from the resulting
recovered slice are then used for comparison with the original
slice. The data recovery process is illustrated for one slice in
Figure\,\ref{fig:illustrated_hit}.

\begin{figure}[!tbp]%
  \centering%
  \begin{subfigure}[t]{0.48\textwidth}%
    \includegraphics[width=\textwidth]{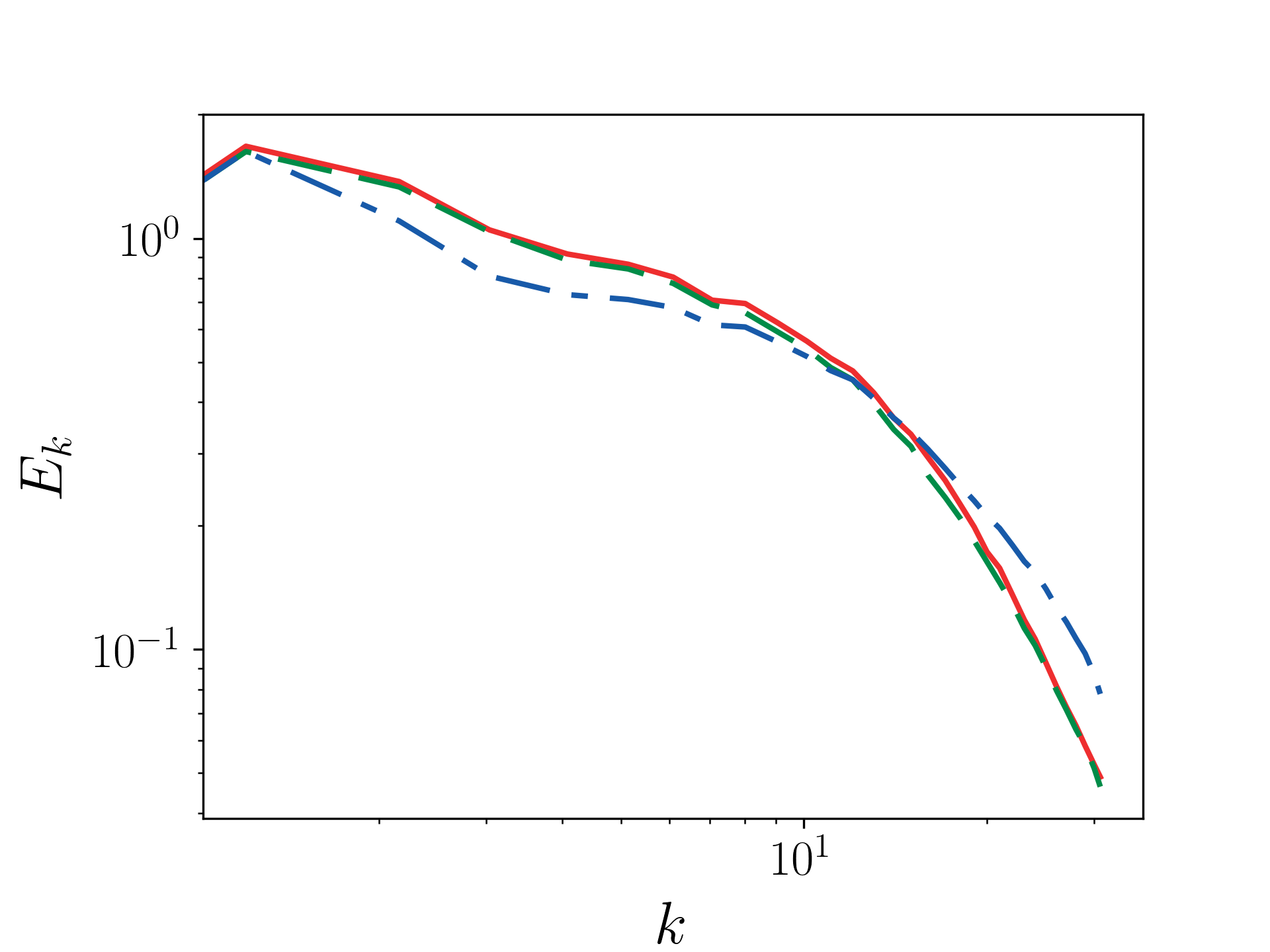}%
    \caption{Average energy spectrum.}\label{fig:spectra}%
  \end{subfigure}%
  \hfill%
  \begin{subfigure}[t]{0.48\textwidth}%
    \includegraphics[width=\textwidth]{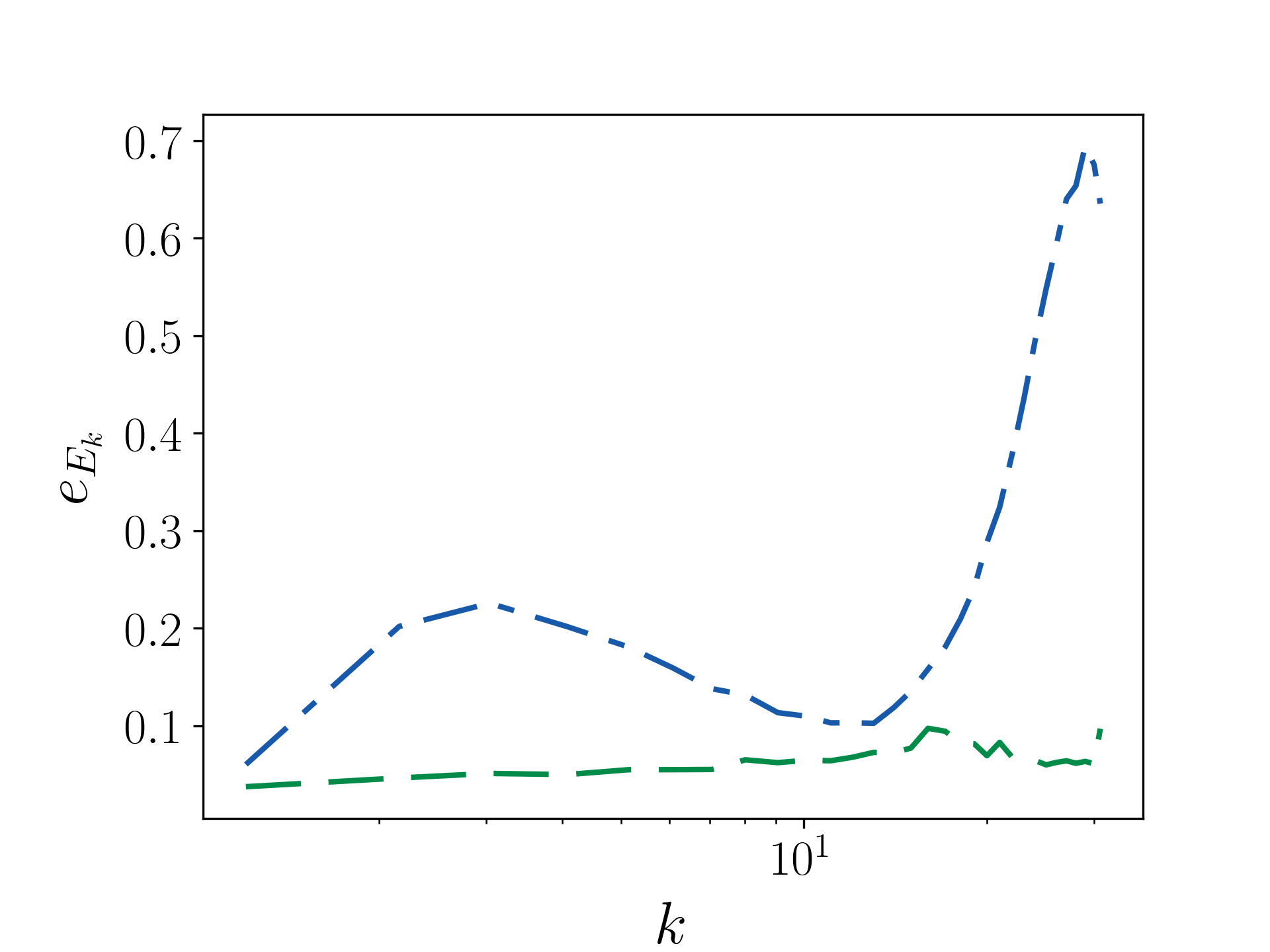}%
    \caption{Normalized error, $e_{E_k} = \frac{|E^h_k - E_k|}{E_k}$.}\label{fig:error_spectra}%
  \end{subfigure}%
  \caption{Energy spectrum and error as a function of wavenumber $k$. Solid red: original data; dashed green: deep convolutional neural network; dot-dashed blue: \gls{gpr}.}\label{fig:spec}%
\end{figure}

The average error in $u'$ and $\lambda$ for all the reconstructions is
approximately three times larger for the \gls{gpr} process compared to
the deep convolutional neural network. For all the slices, individual
energy spectra are calculated for the original data and the data
recovered through the deep convolutional neural network and
\gls{gpr}. The average energy spectrum is presented in
Figure\,\ref{fig:spectra}. The average error from the \gls{gpr}
reconstruction increases at high wavenumbers, indicating that it not
able to accurately capture the smallest scales of turbulence,
Figure\,\ref{fig:error_spectra}. This behavior is not exhibited with
the deep neural network reconstruction, and the error increases
slightly as a function of wavenumber.

\begin{figure}[!tbp]%
  \centering%
  \begin{subfigure}[t]{0.48\textwidth}%
    \includegraphics[width=\textwidth]{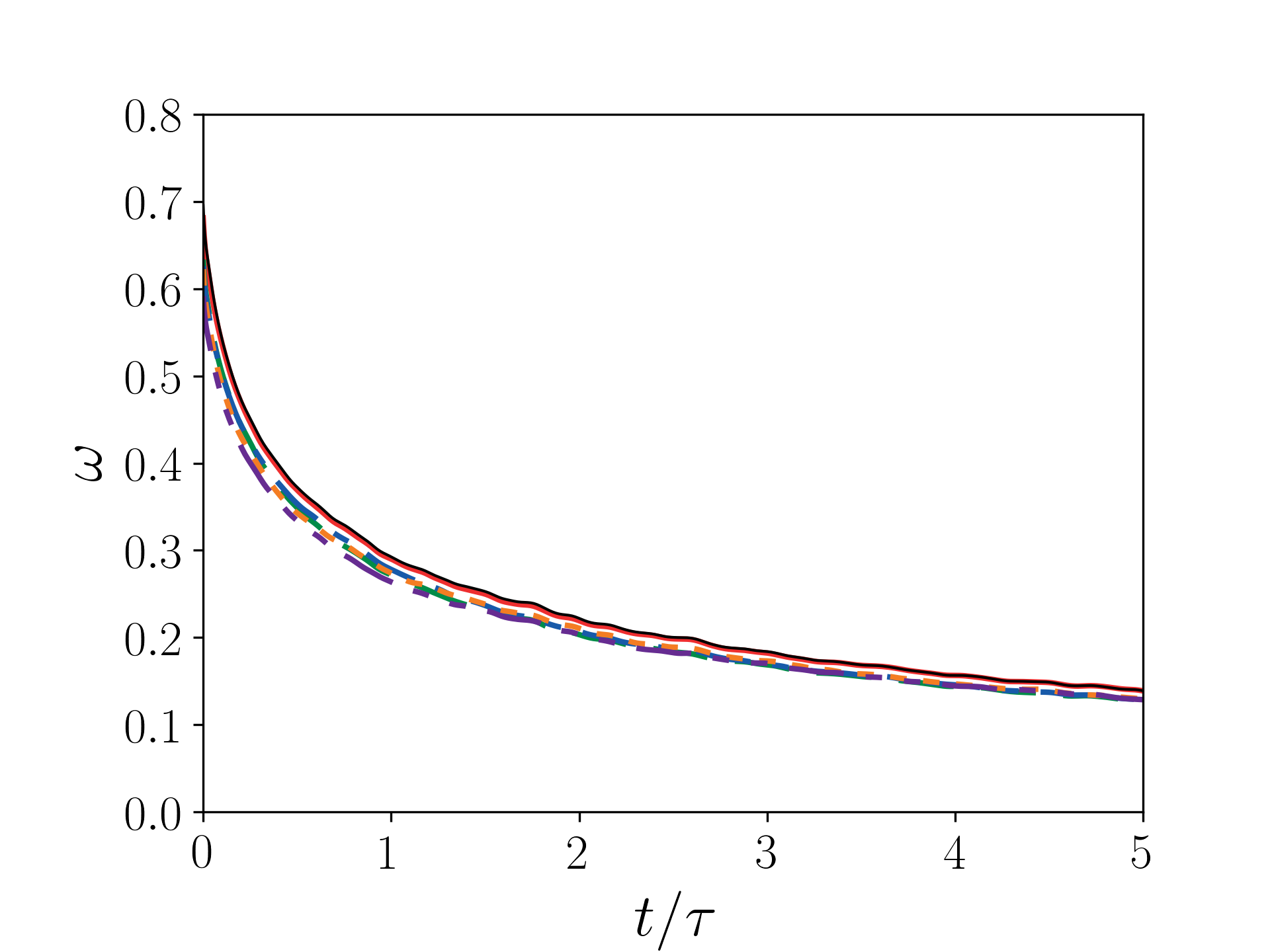}%
    \caption{Deep convolutional neural network.}\label{fig:enstrophy_dl}%
  \end{subfigure}%
  \hfill%
  \begin{subfigure}[t]{0.48\textwidth}%
    \includegraphics[width=\textwidth]{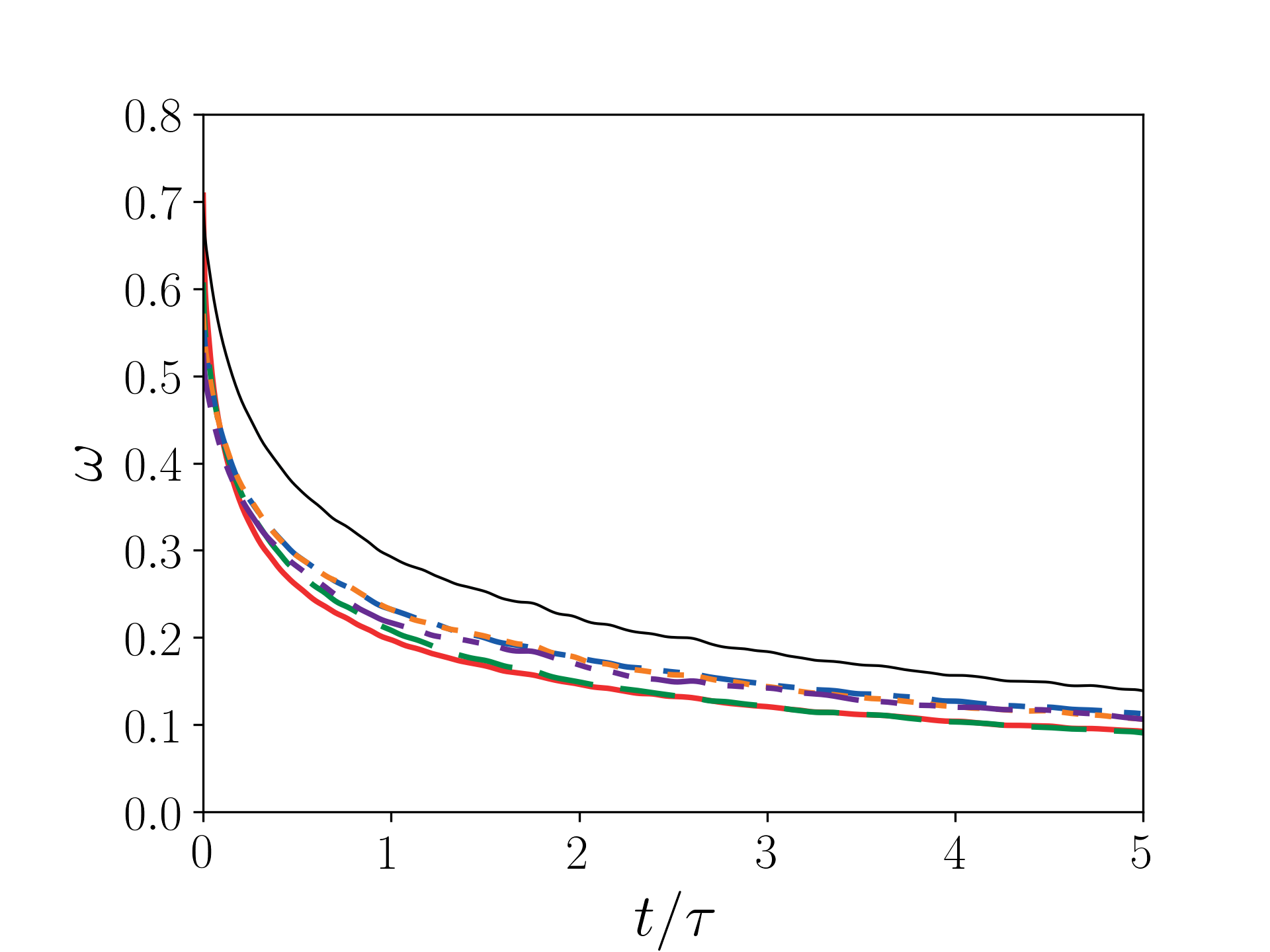}%
    \caption{\gls{gpr}.}\label{fig:enstrophy_gp}%
  \end{subfigure}%
  \caption{Normalized enstrophy as a function of time where 25\% of
    the original data is missing. Solid black: original data; solid
    red: $L_m=0.74\lambda$; dashed green: $L_m=1.48\lambda$;
    dot-dashed blue: $L_m=2.97\lambda$; dotted orange:
    $L_m=5.94\lambda$; dot-dot-dashed purple:
    $L_m=11.87\lambda$.}\label{fig:hit_enstrophy}%
\end{figure}

Velocities in slices of the original and recovered data are used as
initial conditions for two-dimensional decay simulations. In these
simulations performed in PeleC, the initial conditions for the
velocity fields are taken from the (i) original data, (ii)
recovered data from the deep convolutional neural network, and (iii)
recovered data from the \gls{gpr}. The final time for the simulations
is $5\tau$, where $\tau =
\nicefrac{\lambda}{u'}$. Figure\,\ref{fig:hit_enstrophy} illustrates
the decay of normalized enstrophy,
$\omega = \frac{\lambda^2}{u'^2 V}\int_V \left( \nabla \times u
\right)^2 \mathrm{d}V$, for simulations where $f=25\%$ and
$L_m \in [0.74\lambda, 11.87\lambda]$, or $3.125\%$ to $50\%$ of the
domain length. \gls{gpr} reconstruction exhibits a significantly
different enstrophy decay from the original data,
Figure\,\ref{fig:enstrophy_gp}. For the deep convolutional neural
network, the enstrophy decay is well captured at all mask sizes, with
slightly less accuracy for large $L_m$,
Figure\,\ref{fig:enstrophy_dl}. Across the range of mask sizes, the
normalized error at $t=1\tau$ in kinetic energy and enstrophy is less than
10\% for the deep convolutional neural network and around 20\% for the
\gls{gpr} reconstruction. These differences in reconstruction
procedures are attributed to the deep convolutional neural network's
ability to preserve the energy spectra and accurately represent all
the length scales.

\section{Conclusion}\label{sec:ccl}

This work evaluated the use of deep convolutional neural networks for
data recovery of fluid flows in the context of data loss because of
hardware or software faults. The method proposed here leverages an
encoder-decoder deep convolutional neural network to transform a fixed
input to a recovered output using only deteriorated data and
eschewing a training database. Comparisons were performed with
\acrlong{gpr}, a standard data recovery algorithm often referred to as
Kriging. Data recovery was performed on two different canonical flow
configurations: laminar flow around a cylinder and homogeneous
isotropic turbulence. For data recovery of the laminar flow around a
cylinder, results indicate similar performance between the proposed
method and \gls{gpr} across a wide range of mask sizes. For homogeneous
isotropic turbulence, data recovery through the deep convolutional
neural network exhibits an error in relevant turbulent quantities
approximately three times smaller than that for the \gls{gpr}. Forward
simulations using recovered data illustrate that the enstrophy decay
is accurately captured using the deep convolutional neural network
approach.

We emphasize that the work presented here is not necessarily beholden
to the specific inpainting technique used. The deep learning community
has developed many different methods for image inpainting that can be
used for data recovery, and it is expected that state-of-the-art
methodologies would perform comparably well. Rather, the use of deep
image priors as first proposed by Ulyanov et al.~\cite{Ulyanov2017}
and investigated here provides a convenient framework to perform data
recovery of fluid flows because it does not require pretraining the neural
network to construct image priors for different flows. It is therefore
agnostic to the specific flow configuration, and the same framework can
be used for very different flows. This technique, however, does
necessitate the solution of an optimization problem for each data
recovery task. Future work will investigate alleviating this through
partial pretraining, transfer learning, and perceptual loss
functions.

This work --- including data sets, demonstration notebooks, analysis
scripts, and figures --- can be publicly accessed at the project's GitHub
page.\footnote{\url{https://github.com/NREL/deep-image-prior-cfd}} Traditional
machine learning algorithms were implemented through
scikit-learn~\cite{Pedregosa2011} and the deep learning algorithms
through PyTorch~\cite{Paszke2017}.

\section*{Acknowledgments}
This work was authored by the National Renewable Energy Laboratory, operated by Alliance for Sustainable Energy, LLC, for the U.S. Department of Energy (DOE) under Contract No. DE-AC36-08GO28308. Funding provided by U.S. Department of Energy Office of Science and National Nuclear Security Administration. The views expressed in the article do not necessarily represent the views of the DOE or the U.S. Government. The U.S. Government retains and the publisher, by accepting the article for publication, acknowledges that the U.S. Government retains a nonexclusive, paid-up, irrevocable, worldwide license to publish or reproduce the published form of this work, or allow others to do so, for U.S. Government purposes.

This research was supported by the Exascale Computing Project (ECP), Project Number: 17-SC-20-SC, a collaborative effort of two DOE organizations -- the Office of Science and the National Nuclear Security Administration -- responsible for the planning and preparation of a capable exascale ecosystem -- including software, applications, hardware, advanced system engineering, and early testbed platforms -- to support the nation's exascale computing imperative.

\section*{References}
\bibliography{library}

\end{document}